\documentclass[conference]{IEEEtran}
\IEEEoverridecommandlockouts
\usepackage{cite}
\usepackage{amsmath,amssymb,amsfonts}
\usepackage{algorithmic}
\usepackage{graphicx}
\usepackage{textcomp}
\usepackage{xcolor}
\usepackage{siunitx}

\DeclareSIUnit\nauticalmile{NM}

\def\BibTeX{{\rm B\kern-.05em{\sc i\kern-.025em b}\kern-.08em
    T\kern-.1667em\lower.7ex\hbox{E}\kern-.125emX}}
    
\begin{document}

\title{A resource management approach for concurrent operation of RF functionalities
\thanks{This project has received funding from the European Union's Preparatory Action on Defence Research under grant agreement No 882407 [CROWN].}
}

\author{\IEEEauthorblockN{Pascal Marquardt, Sebastian Durst, Kilian Barth and Tobias Müller}
\IEEEauthorblockA{\textit{Cognitive Methods Department} \\
\textit{Fraunhofer FHR}\\
Wachtberg, Germany \\
pascal.marquardt@fhr.fraunhofer.de}
}

\maketitle

\begin{abstract}
Future multifunction RF systems will be able to not only perform various different radar, communication and electronic warfare functionalities but also to perform them simultaneously on the same aperture.
This ability of concurrent operations requires new, cognitive approaches of resource management compared to classical methods.
This paper presents such a new approach using a combination of quality of service based resource management and Monte Carlo tree search.
\end{abstract}


\section{Introduction}
Modern and future Multifunctional RF-Systems (MFRFS) will be capable of performing not only radar functionalities but also communication and electronic warfare (EW) functions.
New technological approaches even foresee to perform these functionalities simultaneously on a single aperture \cite{Heras2022}.
A (far from complete) list of such functionalities comprises for radar operations Air-to-Air (A/A), Air-to-Ground (A/G), and Air-to-Sea (A/S) volume \& surface search, tracking, reconnaissance with Spot/Strip Synthetic Aperture Radar (SAR) and maritime Inverse Synthetic Aperture Radar (ISAR) functions, for EW operation Electronic Support (ES), Electronic Intelligence (ELINT), Radar Warning, and Electronic Counter Measures (ECM) functions, and for data link in network-centric operation Line of Sight (LoS) high data rate links, Beyond Line of Sight (BLoS) data links, as well as Command \& Control Communications.

The sheer mass of different functionalities of an MFRFS combined with the ability of concurrent operations creates the need for new concepts of cognitive resource management. 
This paper introduces a concept for quality of service based resource management extended by the capability to cope with concurrent modes. A main focus lies in the transfer of theoretical considerations into real-world applications.
There are two steps necessary to do this, first the development of models to evaluate the quality of a certain task under given (environmental) conditions and resource constraints.
And second, the design of a method to decide if it is useful to perform two tasks concurrently given that they can be executed in several different configurations.
The introduced methods are validated using a software simulation developed at Fraunhofer FHR.

The paper is structured as follows: Section~\ref{sec:qram} gives a short introduction to the general Q-RAM and describes how realistic quality and utility functions can be defined.
Section~\ref{sec:concurrent} describes the adaptation of the framework to cope with concurrent operations.
Section~\ref{sec:results} evaluates the performance of the proposed methods compared to the standard mode without any concurrency.
And section~\ref{sec:conclusion} gives a short conclusion of the paper.

\section{Quality of service based resource management}\label{sec:qram}
\subsection{General framework}
This section briefly introduces the quality of service based resource allocation model (Q-RAM) and its classical solution approach (cf.~\cite{Ghosh2006}).

Q-RAM assigns limited radar \emph{resources}
in an optimal manner to the tasks the system has to perform.
This is done by selecting \emph{operational parameters}
or task configurations.
These configurations are evaluated with respect to certain performance measures, i.e.\ \emph{qualities}, which are also influenced by 
\emph{environmental conditions}.
The quality can be used to implement hard requirements for specific functions of a system and additionally should be defined such that they are interpretable by a human operator.
To compare different qualities, encode mission goals and task priorities, a scalar value called \emph{utility} is associated with a task configuration's qualities and the existing environmental conditions.
Q-RAM then tries to maximize the sum of all utilities.

Mathematically, this can be formulated as follows (taken from \cite{Durst2021}).
Let $\{\tau_1,\ldots,\tau_n\}$ be a set of radar tasks and let there be $k$ types of resources with resource bounds $R_1,\ldots,R_k$.
Associated with each task $\tau_i$ are
\begin{itemize}
	\item a discrete operational space $\Phi_i$, i.e.\ a discrete space of feasible task configurations,
	\item a function $g_i\colon\thinspace \Phi_i\rightarrow\mathbb{R}^k$ mapping task configurations to their resource requirements,
	\item a quality space $Q_i$ and
	an environment space $E_i$,
	\item a map  $f_i\colon\thinspace \Phi_i\times E_i\rightarrow Q_i$ associating a quality level to a configuration-environment-pair and
	\item a quality-based utility function $\widetilde{u}_i\colon\thinspace Q_i\times E_i\rightarrow\mathbb{R}$.	
\end{itemize}
We define $u_i\colon\thinspace \Phi_i\times E_i\rightarrow\mathbb{R}$ via
$u_i(\phi, e) := \widetilde{u}_i(f_i(\phi, e), e)$
and the \emph{system utility} $u$ for chosen configurations
$\phi = (\phi_1,\ldots,\phi_n) \in \Phi := \Phi_1\times\cdots\Phi_n$
under environmental conditions
$e=(e_1,\ldots,e_n) \in E := E_1\times\cdots E_n$ as
$
u(\phi, e) = \sum_{i=1}^{n} u_i(\phi_i, e_i).
$
Now for fixed environmental data $e\in E$, the aim is to optimize global system utility while respecting resource bounds, i.e.\ we have 
the following optimization problem:
\begin{align}
	\begin{split}
	\max_{\phi = (\phi_1,\ldots,\phi_n)}& u(\phi, e)\\
	\textrm{s.t. } \forall j=1,\ldots,k\;\;& \sum_{i=1}^n \big(g_i(\phi_i)\big)_j \leq R_j.
	\end{split}
\end{align}

A solution to the Q-RAM problem is proposed in \cite{Rajkumar1997, Lee1998, Lee1999, Ghosh2004, Ghosh2006} and will be briefly outlined in the following.
If there are multiple types of resources $R_1,\ldots,R_k$, a so-called \emph{compound resource} is used, i.e.\ a function
$h\colon\thinspace\mathbb{R}^k\rightarrow\mathbb{R}$ mapping a resource vector to a scalar measure of resource requirements.
On a per task basis, all possible task configurations are generated and evaluated. This yields an embedding from the space of task configurations into resource-utility-space. A convex hull operation is used to determine the subset of configurations maximizing utility for fixed resource levels.
A global optimizer then iteratively allocates resources to the task offering the best utility-to-resource-ratio provided sufficient resources are available.
After the resource allocation step, the resulting tasks are placed on the timeline by a scheduler.

\subsection{Performance models}
The use of the Q-RAM framework requires functions describing the quality, utility and resource usage of a task configuration under given environmental conditions (which include system and mission parameters).
Especially when transferring the Q-RAM based resource management to a real-world MFRFS the definition of such functions turns out to be a rather complex problem.

In this article we introduce the term \textit{performance model} to describe the assembly of the quality of a task using certain resources as well as the utility (potentially from a mission perspective).
Thus, the model summarises all calculations required by the Q-RAM model for a specific task.

Hence, a performance model is defined to be a set of mathematical functions to evaluate the expected or actual benefit (i.e.\ quality and utility) and resource requirements of a task in a given configuration taking into account environmental conditions.
A performance model can consist of parametrised functions to allow for mission specific adaptations and control.

Input dimensions of the model are operational parameters, i.e.\ steerable parameters that are variables in regard of the resource allocation, and environmental parameters.
Environmental data is all task related data that impacts the performance but is outside the control of the resource manager, e.g.\ target related data for a tracking task, distance to the other party for a communication task or information on clutter and jammers, but also mission specific considerations.
The actual input to the model then is a list of possible task configurations and the current environmental conditions.

The model then outputs resource requirements and expected quality and utility.

The design of such performance model can be summarised in four steps.
These will be illustrated in the following using the example of a radar tracking task.

\subsubsection{Qualitative description}
The first step is to identify the goal of the mode.
For a tracking task this might be "Localise a known target over a given time".
From that, one can derive the requirements of the mode, e.g.\ "Maintain a target track with a given accuracy".
This leads to the quality measures of the mode.
For the example of tracking this is most obviously the tracking error.
Note that in most cases there will be more than one quality measure that have to be taken into account.

\subsubsection{Determining relevant parameters}
In the second step relevant parameters will be identified.
We distinguish four different types of parameters:
\begin{itemize}
	\item \textbf{Operational parameters:} Parameters that are variable during resource allocation and that are directly controlled by the resource manager (e.g.\ pulse length, number of integrated pulses, number of antenna elements).
	\item \textbf{Environmental parameters (depending on target / task):} Parameters that are not controlled by the resource manager depending on the environment around the system (e.g.\ clutter, target dynamics). 
	\item \textbf{Environmental parameters (system parameters):} Parameters that are not controlled by the resource manager coming from the system (e.g.\ maximum radiating power, receiver temperature).
	\item \textbf{Environmental parameters (mission-defined):} Parameters that are not controlled by the resource manager coming from mission control (e.g.\ desired false alarm rate, volume of interest).
\end{itemize}

\subsubsection{Mathematical formulation of quality and resource requirements}
In the third step the quality measures and resource requirements can now be modelled mathematically.
Taking the example further, the quality associated with a configuration of a radar tracking task under given environmental conditions is defined as the inverse of the expected track error after performing the task in this configuration:
\begin{equation}
	q = \frac{1}{e_t},
\end{equation}
where $e_t$ is the expected track error.

A task configuration $c=\left(n_{az},n_{el},\mathrm{PRF} ,n_p,\tau,B,\lambda\right)$\footnote{$n_{az}$ and $n_{el}$ are the number of antenna elements in azimuth and elevation, respectively; $PRF$ is the pulse repetition frequency; $n_p$ is the number of pulses to be integrated; $\tau$ is the pulse length; $B$ is the bandwidth; $\lambda$ is the wavelength associated with centre frequency} for tracking a target at range $R$ has the following resource requirements:
\begin{itemize}
	\item Number of antenna elements used 
	\begin{equation}
		n_{tot}=n_{az} n_{el}
	\end{equation}
	\item Task duration $T_{task}$, i.e.\ the time the chosen elements are in use: 
	\begin{equation}
		T_{task} = \frac{n_p-1}{PRF} + \tau + \frac{2R}{c}
	\end{equation}
\end{itemize}
Note that the power budget does not have to be considered as all tasks are enforced to respect the system's duty cycle.

\subsubsection{Formulation of utility function}
In the fourth step a utility function for the mode has to be defined.
Whereas the quality functions describe objective measures, the utility is more subjective and depending on target priorities and mission goals.
An example (taken from \cite{Ghosh2006}) of a utility function for a track update task is
\begin{equation}
	u=w\left(1-e^{-\beta q}\right),
\end{equation}
where $w$ and $\beta$ are still to be defined weight functions that can depend on environmental factors like target dynamics, target priorities and mission goals.
An example of an appropriate $w$ is
\begin{equation}
	w=K_t\left(\frac{v}{R+K_R}\right)
\end{equation}
where $v$ is the radial velocity of the target, $K_t$ a priority depending on the target type and $K_R$ a constant.

\section{Q-RAM for concurrent operations}
The resource manager shall be capable to optimise different concurrent working modes.
As described in Section~\ref{sec:qram}, the regular operation uses the Q-RAM algorithm to allocate resources.
For concurrent modes an enhanced approach based on the same framework will be used.
Before going into details of that framework there will be a short section on how the term concurrency is used in this paper.

\subsection{Concurrent operation modes}\label{sec:concurrent}
There are several different ways to define and understand concurrency in an MFRFS.
Therefore, this section gives a short overview of the concurrent operation modes that were used for our investigation.

\subsubsection{Interleaved mode}\label{subsec:interleaved}
This mode combines radar, electronic warfare and communication functions in a coordinated (non-concurrent) operation.
RF tasks are interleaved in time on a beam level, using the whole aperture.
In future investigations this mode might be expanded to interleaving on a pulse level.

\subsubsection{Multifunction mode}\label{subsec:multifunction}
This mode combines multiple tasks in a single, specifically designed waveform using the whole aperture.
For example, radar and communication tasks can be combined into a single task using a multifunction waveform for a more efficient use of the timeline.
A second example could be the combination of radar and electronic warfare enabling the system to simultaneously track and jam an approaching missile.

\subsubsection{Multioperation mode}\label{subsec:multioperation}
This mode performs multiple RF tasks, possibly of different functions (e.g. radar and electronic warfare) and in different directions, concurrently by using a different subarray of the aperture for each task.
Transmission is concurrent and subject to hardware limitations like isolation of subarrays.
Transmit and receive might be concurrent on the different sub-apertures in future investigations.

\subsection{Monte Carlo Tree Search for concurrent modes}
In its standard implementation the Q-RAM algorithm requires an iterative global optimisation process for resource allocation.
When dealing with concurrency, this is not suitable anymore since the decision whether a task should be executed solely or in combination with other tasks has to be made in advance and is not changeable during the optimisation.

A simple solution would be to only consider fixed (and not all possible) combinations of tasks to be executed concurrently.
The goal of the proposed algorithm is to find and evaluate among all possible combinations the most promising ones.
The set of all combinations forms a tree in the following way.
Starting from the root a branch is formed for each task and its combination with other tasks.
This process is done recursively where every task is only allowed to appear once (solely or in combination) inside each path from the root to a leaf.
%
%
\begin{figure}
\centering
\includegraphics[width=0.85\columnwidth]{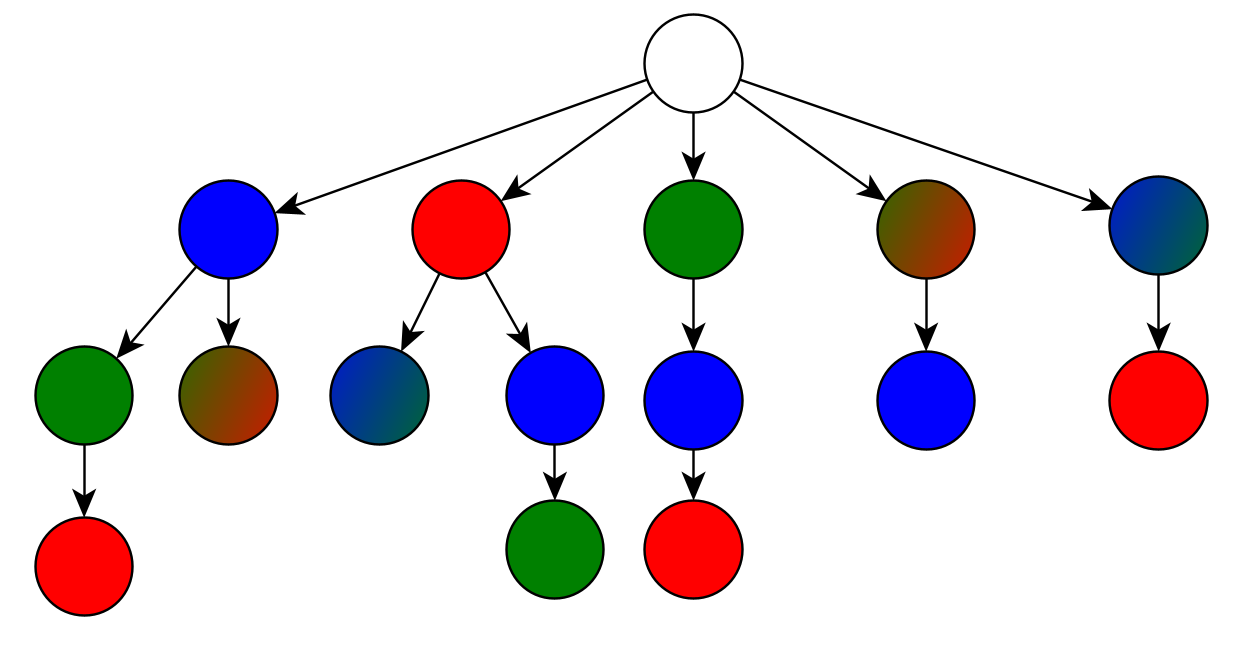}
\caption{Tree approach for best utility. A single colour corresponds to a single mode configuration. Two colours refer to a combination of two tasks.}
\label{fig:tree}
\end{figure}
The branching might be bounded by technical or tactical restrictions like suppressing beams into the same direction working in the same frequency domain.

For each leaf of this tree an individual Q-RAM optimisation has to be calculated.
Afterwards the path with the highest utility is chosen for scheduling.
For an accurate optimisation proper performance models for combined tasks are required.
 

In Fig.~\ref{fig:tree} an example tree is depicted.
A path from the white root to a leaf represents an allowed subset, e.g.\ blue, green and red or blue and green/red.

In a realistic operational setting, the number of tasks and combinations quickly becomes too large to perform Q-RAM for all of them.
To deal with this problem, Monte Carlo Tree Search (MCTS) is implemented \cite{Browne2012}.
The details of this development and implementation will be the subject of an upcoming publication by the same authors.

%

\section{Experimental results}\label{sec:results}
\subsection{Simulation environment}
As part of the basic funding by the German Ministry of Defence, a powerful simulator for phased array radars with electronic beam steering was developed at Fraunhofer FHR.
This simulator enables real-time analyses of radar systems in different frequency ranges, rotating or static, with arbitrary antenna patterns and search strategies.
The simulation includes the most important functions, such as search, tracking, target classification, resource management and data fusion across multiple sensors.
The simulation also enables the analysis of threat trajectories, taking into account aspect angle-dependent RCS values.

\begin{figure}
\centering
\includegraphics[width=0.85\columnwidth]{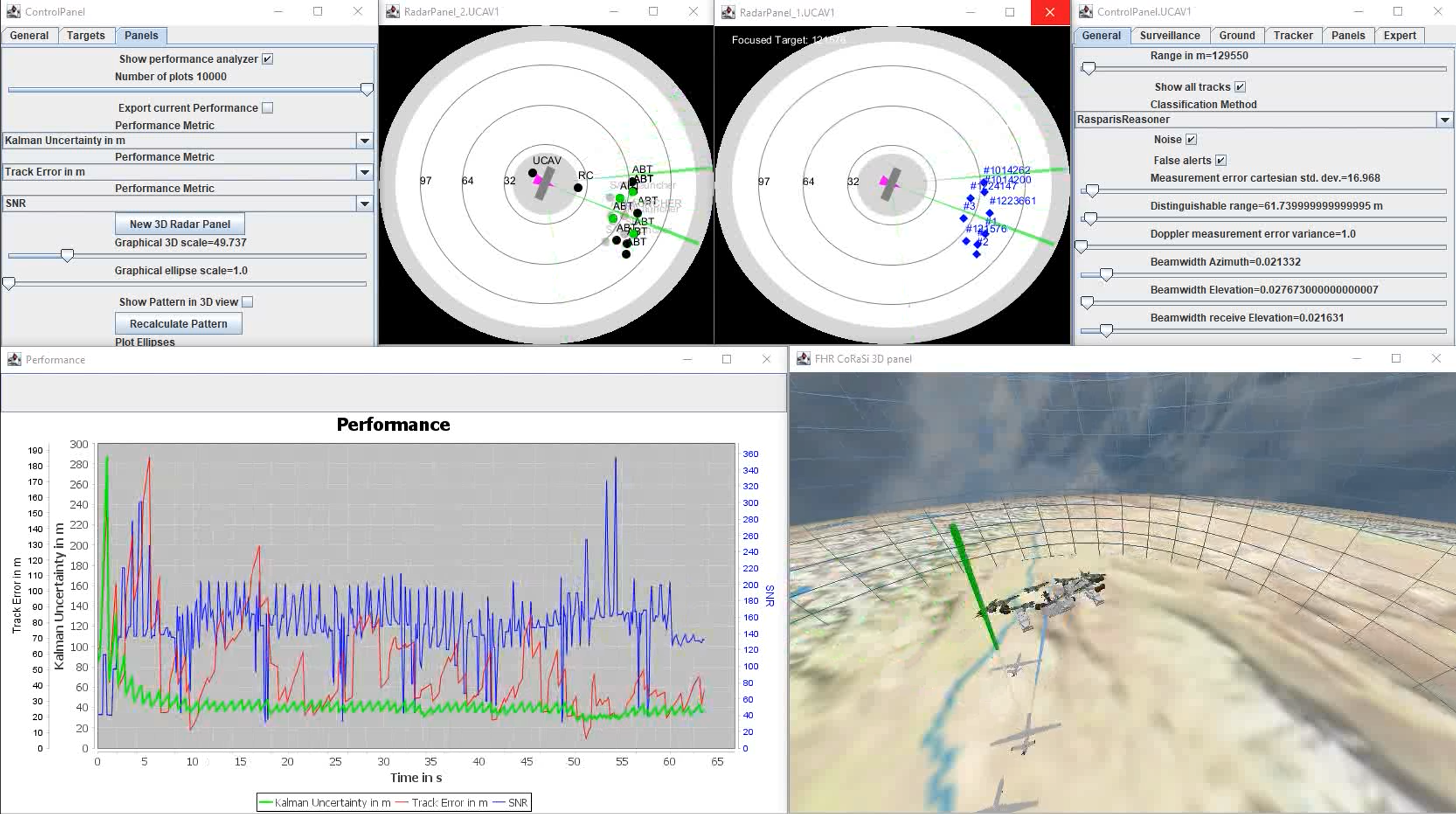}
\caption{CoRaSi user interface.}\label{fig:corasi}
\end{figure}

The Cognitive Radar Simulator \textit{CoRaSi} is a Java application that can be used for various purposes.
For a better usability, CoRaSi comes with a 3D-GUI (Fig.~\ref{fig:corasi}).
There it is possible to see ground-truth as well as the sensor's view of the scenery.
Several sliders and drop-down menus allow the user to change sensor settings and scenario parameters while the simulation is running.
Through a performance graph different performance metrics can be evaluated.

\subsection{Scenario}
For the evaluation a complex scenario has been chosen\footnote{The scenario has been derived from the CONOPS of the CROWN project.} with the goal of destroying a ballistic missile launcher.
The launcher is located \SI{300}{\nauticalmile} behind the forward line of allied troops.
Its estimated position is known beforehand by intelligence within an area of \SI{25}{\square\kilo\meter}.
The scenario also includes enemy air patrol and surface-to-air missiles (SAMs).

In a phase of the scenario the unmanned combat aerial vehicles (UCAVs) and a remote carrier (RC) are going to fly to the region to detect the actual launcher position.
For the identification of the exact position it is necessary to carry out a SAR image of the scenery with one of the UCAVs and communicate it to some ground station for classification.
The enemy is defending the ballistic missile launcher using Medium Range SAM (MRSAM).
The resource management for the demonstration will be considered for one of the UCAVs called UCAV 1.

\begin{figure}
\centering
\includegraphics[width=0.85\columnwidth]{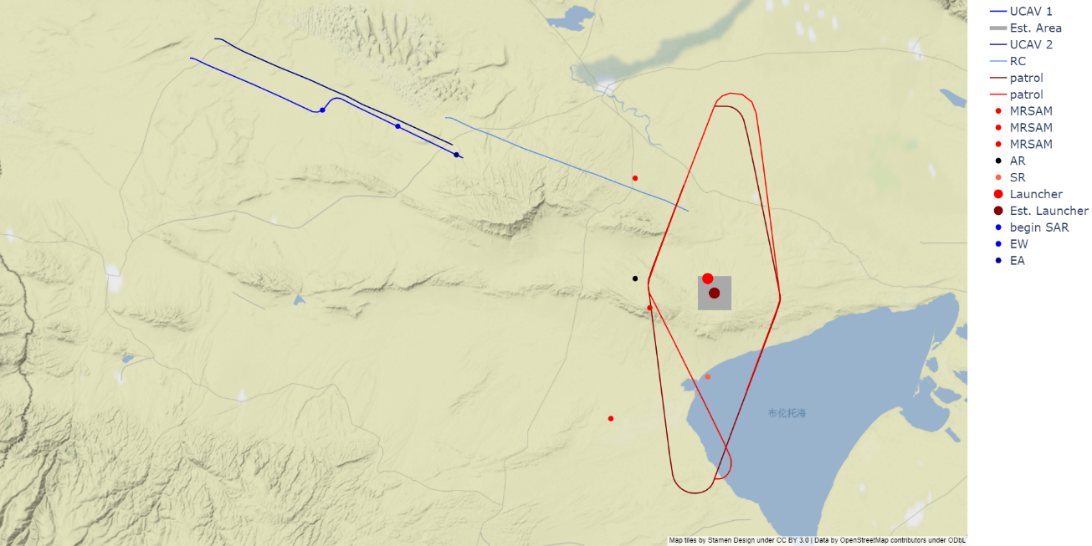}
\caption{The scenario in top view with all trajectories and positions. The resource management takes place in UCAV 1.}\label{fig:scenario}
\end{figure}

Fig.~\ref{fig:scenario} depicts all trajectories and positions of all necessary elements of the scenario.
Friendly vehicles are shown in blue colours and enemy vehicles in red.
Moving planes are plotted with lines and the position of standing trucks are marked by circles.
Blue circles on the trajectory of UCAV 1 mark planned (SAR) or unplanned (EW, EA) events.

In this scenario UCAV 1 has to perform ten different RF modes: five for radar (A/A surveillance, A/A tracking, A/A High Range Resolution Profiles, A/G Stripmap SAR, A/G Wide Area Scan Ground Moving Target Indication), three for EW (Radar Warning Receiver, Surveillance ESM Defensive, Electronic Attack), two for communication (data link, SAR image LoS communication).

\begin{figure}
\centering
\includegraphics[width=0.85\columnwidth]{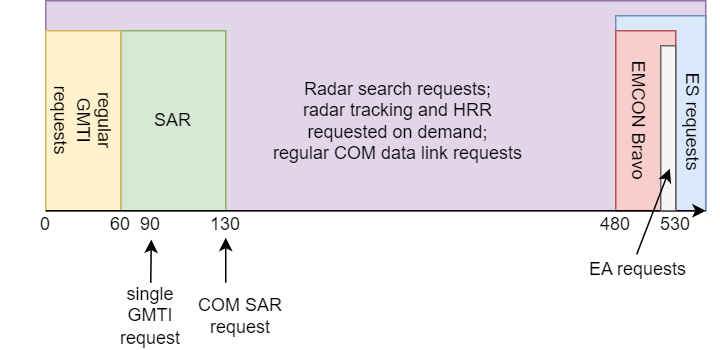}
\caption{Graphical representation of the storyboard.}\label{fig:storyboard}
\end{figure}

The simulated phase of the scenario lasts about \SI{550}{\s} and consists of several different request of tasks during its runtime (see Fig.~\ref{fig:storyboard}).
A task request doesn't necessarily mean that the task is scheduled to the antenna timeline but that there is a demand for it, e.g.\ coming from the mission or an operator.
The resource management decides on the basis of the other requests and mission goals which tasks in which configurations yield the best contribution to the total system utility.

For the storyboard there are some tasks that are requested on a regular base like surveillance or tracking but also the communication data link.
Other tasks like the GMTI task or the SAR task are requested at specific times.
The GMTI tasks at the beginning of the scenario are highly resource demanding and thus a challenge for the resource manager.
The SAR task starting at around \SI{60}{\s} usually blocks the antenna for other tasks in a standard operation.
At around \SI{480}{\s} emission control (EMCON) level Bravo is enforced, i.e.\ RF transmissions are only allowed for communication – radar transmissions are forbidden.

\subsection{Results}
For the evaluation of the performance all modes were simulated with \num{25} randomised Monte Carlo simulations each.
The general scenario stayed the same but was started with slightly different timings and starting points on the flight trajectories to create some statistical variation.
The three concurrent operation modes described in section~\ref{sec:concurrent} are compared to the standard mode where no concurrent operation is possible.

The major impact of the concurrency can be seen in the track error.
This is why we will concentrate on this performance metric.
A low tracking error indicates that the mission goal of reconnaissance of the scenery can be done well.

\begin{figure}
\centering
\includegraphics[width=0.8\columnwidth]{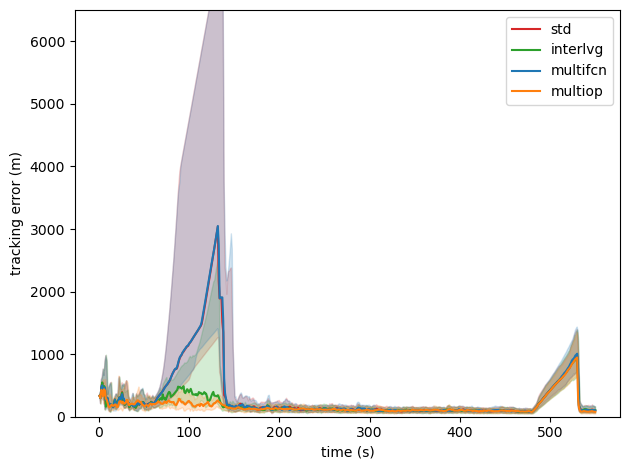}
\caption{Track error over time.}\label{fig:track_error}
\end{figure}

Fig.~\ref{fig:track_error} shows the the mean track error over time for all simulation runs and operation modes over time.
There are two major time frames where the performance of the modes are distinguishable.

\begin{figure}
\centering
\includegraphics[width=0.8\columnwidth]{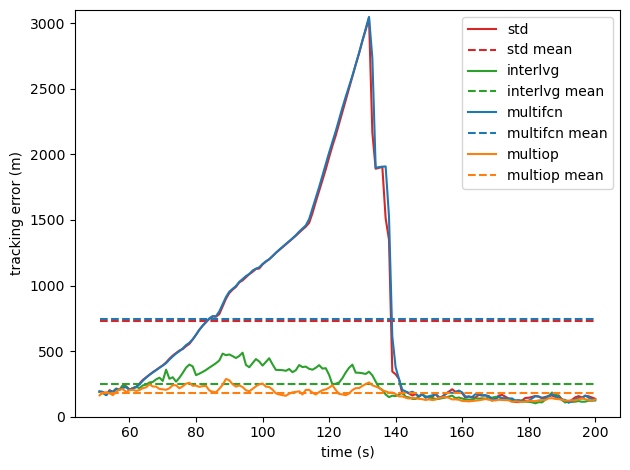}
\caption{Track error while performing stripmap SAR.}\label{fig:track_error_SAR}
\end{figure}

The first time frame of interest is roughly between \SI{60}{\s} and \SI{130}{\s}.
In this interval, the system performs a stripmap SAR task, independent of the mode in question.
The standard and multifunction mode do not have the ability to actively track targets in this time frame.
The interleaving and multioperation mode on the other hand can still track by utilising their ability for concurrent operations.
The track error for interleaved and multioperation mode increases with time.
Compared to the mean in this time interval, the track error of the multioperation mode increases by roughly \SI{100}{\m} while the track error of the interleaving mode doubles to \SI{500}{\m}, see Fig.~\ref{fig:track_error_SAR}.

The track error of the standard and multifunction mode on the other hand increases to more than \SI{3000}{\m} (Fig.~\ref{fig:track_error_SAR}).
Looking at the third quantile (coloured areas) in Fig.~\ref{fig:track_error}, the track error is in the worst case around \SI{6000}{\m}, while the interleaving and multioperation mode are below \SI{1500}{\m} and \SI{500}{\m} respectively.

The second time frame of interest is after the change in EMCON level at the end of the scenario.
All modes stop actively tracking which also leads to an increased track error.
The multioperation mode has a slightly better performance, which is not statistically significant.

\begin{figure}
\centering
\includegraphics[width=0.8\columnwidth]{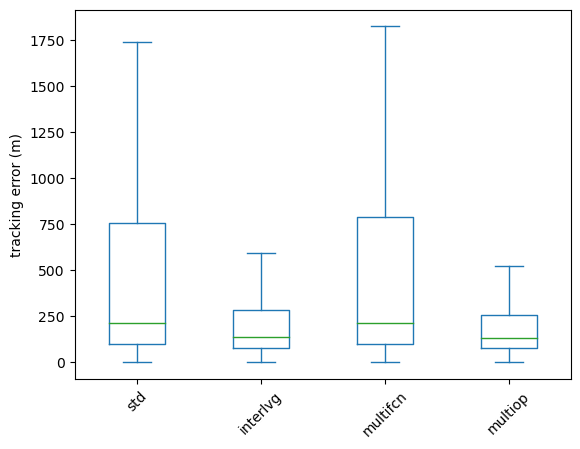}
\caption{Total track error.}\label{fig:track_error_total}
\end{figure}

The total track error depicted in Fig.~\ref{fig:track_error_total}, shows the average track error in the scenario as well as the deviation.
The average of the four modes is relatively similar with a median of roughly \SI{220}{\m} for the standard and multifunction mode and \SI{140}{\m} for the interleaved and multioperation mode.
The third quantiles and standard deviation for the interleaved and multioperation mode on the other hand are below that of the standard and multifunction mode by several hundreds of meters.
This shows that the multifunction and multioperation mode are much more stable and have a track error reduction of \SI{33}{\percent}.

The utility of a task type can be taken as a performance metric as well.
A non-weighted utility of \num{0} means that a task does not satisfy the minimum requirements, while values close to \num{1} mean the function performs to full satisfaction.

\begin{figure}
\centering
\includegraphics[width=0.8\columnwidth]{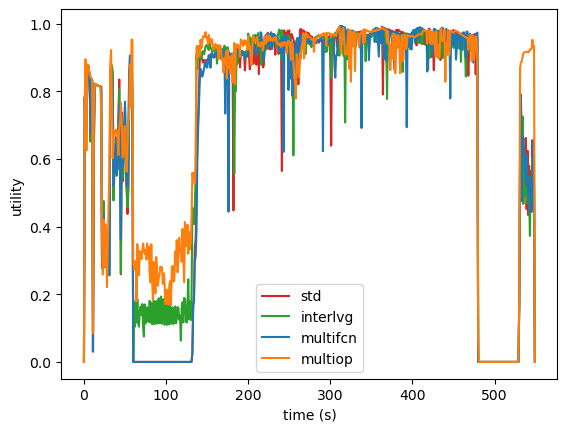}
\caption{Utility of the track update task.}\label{fig:track_utility}
\end{figure}

Fig.~\ref{fig:track_utility} shows that the multioperation mode has, except while conducting the stripmap SAR task, a much lower variance of the utility.
The interleaving mode has also smaller drops with regard to the standard mode.
In terms of average utility, the four modes are nearly the same outside of the stripmap SAR time frame.

\begin{figure}
\centering
\includegraphics[width=0.8\columnwidth]{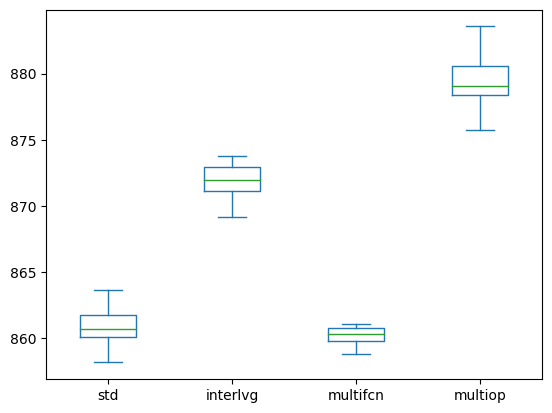}
\caption{Utility of the different modes for the whole scenario.}\label{fig:utility}
\end{figure}

Finally, the utility of the whole scenario duration is depicted in Fig.~\ref{fig:utility}.
The box plots were calculated using the cumulated utilities in the individual Monte-Carlo runs.
It can be seen that multioperation mode outperforms all other modes, with interleaved mode a clear second.
Standard and multifunction mode are comparable with multifunction having a lower variance and more consistent results.
The similarity of these two modes is expected in this setting, as the multifunction mode can only be used in certain geometries.
%

These results show the successful operation of the resource manager.
Resources are fully used (when not restricted by EMCON) and the system achieves an overall high utility whilst balancing different functions.
Mission goals are accomplished in all modes under consideration.
Nevertheless, there are still significant differences in the performance of the modes.

Most strikingly, the tracking performance in multioperation and interleaved mode outperforms multifunction and standard mode by a high margin.
This is mainly due to the fact that tracks can be updated during the long mission-critical SAR operation in the two former modes and thus be kept at a sufficient accuracy, while this is not possible in the two latter modes.

Comparing the different concurrent modes, the seemingly low performance of the multifunction mode stands out.
However, in the scenario under investigation this is not surprising as the multifunction mode is limited to a combined waveform for radar search and communication for concurrent operation and there is only one platform acting as a receiver. 

In summary, the quality of service based resource management enables an efficient concurrent operation which does not only improve utility under heavy load but is also able to equalise the aftereffects more quickly.
Although not being used at times of low loads, the performance of the concurrent modes is considerably better overall in the simulated scenario, which poses a typical mixture of times of high and low demand.
Among the modes investigated, the multioperation mode, i.e.\ sharing the aperture spatially to execute multiple tasks at the same time, performs best.

\section{Conclusion}\label{sec:conclusion}
In this paper we presented a new approach for resource management for concurrent operations of RF functionalities.
It is based on a classical Q-RAM framework but enhanced in two ways.
Using MCTS the algorithm is now capable to efficiently evaluate thousands of configurations in a short period of time.
This is necessary since the ability of the considered MFRFS to split the antenna in several subapertures raises the already high number of possibilities even more.

Furthermore, we presented a blueprint for the realistic definition of quality measures and utility functions needed in Q-RAM.
This is a big step towards bringing the proposed framework onto operating MFRFS in real-world environments.

The results compared to the standard operation modes show that especially the multioperation outperforms all other modes regarding the track error and utility.
This emphasises that our framework is capable to manage the various functionalities of modern and future MFRFS even if they are performed simultaneously.

\section*{Acknowledgment}
The authors would like to thank the other partners of the PADR project "European active electronically scanned array with Combined Radar, cOmmunications, and electronic Warfare fuNctions for military applications" (CROWN) for the fruitful collaboration. Especially Elettronica S.p.A. (Italy), Indra Sistemas S.A. (Spain) and TNO (Netherlands) for their extremely valuable contribution to the technical details regarding the EW, communication and radar modes, and ONERA (France) for contributing a special multifunction waveform.
Furthermore, we are grateful to Saab AB (Sweden), whose mission-specific view on resource management helped to mature our models and algorithms.

\bibliographystyle{IEEEtran}
\bibliography{CROWN}

\end{document}